\documentstyle[11pt]{article}
\newdimen\paperwidth \newdimen\paperlength \newdimen\margin
\newdimen\vmargin
\textwidth=\paperwidth \advance \textwidth by -2\margin
\advance\hoffset by -1truein \advance\hoffset by \margin
\textheight=\paperlength \advance \textheight by -2\vmargin
\advance\voffset by -1truein \advance\voffset by \vmargin
\advance\textheight by -2\baselineskip
\newcommand{\Section}[1]{\section{#1}\setcounter{equation}{0}}
\begin{document}
\begin{titlepage}
\title{ {\bf   The Correlated Block
   Renormalization Group}}\thanks{On leave of absence from Instituto
de Matem\'aticas y F\'{\i}sica Fundamental, C.S.I.C., Serrano 123,
28006-Madrid, Spain. }
\vspace{2cm}    \author{ {\bf Miguel A. Mart\'{\i}n-Delgado}$\dag$,
\mbox{$\:$} {\bf Javier Rodriguez-Laguna}$\ddag$  \  and {\bf Germ\'an
Sierra}$\ast$ \\ \mbox{}    \\ $\dag${\em Departamento de F\'{\i}sica
Te\'orica I}\\ {\em Universidad Complutense.  28040-Madrid, Spain }\\
$\ddag${\em Instituto de Matem\'aticas y F\'{\i}sica Fundamental.
C.S.I.C.}\\ {\em Serrano 123, 28006-Madrid, Spain } \\
$\ast${\em Theoretische Physik - ETH-Honggerberg}\\
 {\em CH-8093, Switzerland } }
\vspace{5cm}
\date{}
\maketitle
\def\baselinestretch{1.3}
\begin{abstract}
We formulate the standard real-space renormalization group method
in a way which takes into account the correlation between blocks.
This is achieved in a dynamical way by means of operators which
reflect the influence on a given block of its neighbours.
We illustrate our method in the example of the tight-binding model
in 1D and 2D for various types of boundary conditions.

\ \

\ \

\ \

\ \

\end{abstract}

\vspace{2cm} PACS numbers: 75.10 Jm, 05.50.+q, 64.60.Ak

\vskip-17.0cm \rightline{UCM/CSIC-95-12} \rightline{{\bf December
1995}} \vskip3in 
\end{titlepage}
\newpage

\section{Introduction}

The initial success of the Real-Space Renormalization Group
 method applied by Wilson
\cite{wilson} to the study of the Kondo problem rised the hope that
this technique could be generalizable to more complicated many-body
problems in Field Theory and Condensed Matter. Various groups of
physicists working on those areas were able to combine Wilson's ideas
together with Kadanoff's concept of block thereby arriving to a
method called Block Renormalization Group (BRG) \cite{drell},\cite{jullien}.

\noindent From a quantitative and sometimes qualitative point of
view, the BRG procedure proved to be not fully reliable particularly
when compared with numerical approaches, such as the Quantum Montecarlo
method, which were developed at the same time. This was one of the
reasons why the BRG methods remained undeveloped during the 80's
until 1992 were a new numerical method called the Density Matrix
Renormalization Group (DMRG) was proposed to overcome the
difficulties of the old techniques \cite{white}, \cite{white-noack}.

 In order to understand the failure of the BRG method Wilson proposed
in 1986 \cite{wilsontalk} to study the problem of a single particle
moving freely in a 1D box. A straightforward application of the  BRG
techniques gives results which are several orders of magnitude off
the exact values. According to White and Noack \cite{white-noack}
this poor performance is due to the fact that the truncation of
states within each block in the BRG method keeps states which do not
have the appropriate boundary conditions (BC). As a matter of fact,
an isolated block has BC's which are different from those it has
when it is immersed into the rest of the lattice. In
\cite{white-noack} two solutions to this problem were proposed which
in turn give rise to two different methods called Combination of
Boundary Conditions (CBC) and the aforementioned DMRG \cite{white}.
In the CBC method one applies different BC's to each block and mix
them up resulting in an accurate performance. However, this method
is difficult to generalize for interacting systems. On the contrary,
the DMRG method has been applied with great success to interacting
systems \cite{white-huse}, \cite{white-scalapino}. In the DMRG
method each block somehow chooses the exact BC it needs. This is
achieved by putting the block in connection with the rest of  the
lattice. In practical terms, one considers a superblock where the
block is embedded, then one finds the exact ground state of the
superblock (called the target state) and constructs a reduced
density matrix $\rho_B$ for the block. This density matrix $\rho_B$
gives the  {\em best posible} representation of the target state on
the block. Finally, one diagonalizes $\rho_B $ and keeps the
eigenstates with largest eigenvalues in the truncation procedure. In
this fashion, the role of the Block Hamiltonian $H_B$ in the BRG method is
played in the DMRG method by the density matrix, which has the
virtue of containing the effect of the neighbourhood on a given
block.

  The DMRG method has been generalized to 2D classical systems in
references \cite{nishino}, \cite{ostlund-rommer}. In this framework
the density matrix turns out to be intimately related to the Baxter's
corner transfer matrix \cite{nishino-okunishi}, which suggests a
pathway for a deeper understanding of the DMRG techniques.
In reference \cite{q-germanyo}, \cite{qbis-germanyo}
 we have introduced the Renormalization
Quantum Group Method (qRG for short) based on quantum groups
techniques and we mentioned that this qRG method has some analogies
with the DMRG.

\noindent All these works show that the DMRG is not only a powerful
computational method but also a source of inspiration for further
works concerning the RG.
For these reasons, it may be worthwhile to explore different options
or alternatives to the DMRG which may be useful in situations where
the DMRG encounters difficulties, as in the case of 2D quantum systems.
The main message of the DMRG is that blocks are correlated. The
implementation of this idea by means of the
density matrix formalism may
be not the unique way to proceed. On the other hand, the
``onion-scheme" a la Wilson adopted by the DMRG, while being one of
the reasons of its spectacular accuracy, imposes certain limitations.

\noindent At this stage it is not clear how fundamental are the
density-matrix formalism or the onion-scheme for a RG method which
takes into account the correlation between blocks.
One can indeed combine the Kadanoff block method with the use of a
density matrix in the process of truncation, as in reference
\cite{dm-germanyo}. More work remains to be done to see wheather there
is a real improvement of the standard BRG method by combining it with
the DMRG as in \cite{dm-germanyo}.
In this paper we want to explore another possibility which is to give
up both the density matrix and the onion-scheme.
With this point of view in mind, it would seem that we should come
pretty close to the standard BRG method, were it not for the enormous
freedom hidden in a Real-Space RG method. This freedom comes from
the separation of the Hamiltonian into an intrablock $H_B$ and an
interblock $H_{BB}$ Hamiltonian.
This is a source of ambiguities which can be
sometimes mitigated with the aide of symmetry arguments, but not
fully eliminated though. This ambiguity shows up specially for
terms in the Hamiltonian acting at the boundaries of the block.
There are no general criteria as to how to include this type of terms
either into the intrablock or into the interblock Hamiltonians, or
into both! For example, in the 1D Ising model in a transverse field
(ITF model), a choice which preserves the selfduality of the model
attributes some self-couplings to the $H_B$ and others to the
interblock $H_{BB}$, and it yields to an exact value of the critical
point and the critical exponent $\nu$ \cite{fpacheco},
\cite{q-germanyo}. The ambiguity in the splitting of $H$ into the
sum $H_B + H_{BB}$ thus affect deeply the truncation procedure itself,
which is based on the diagonalization of $H_B$. Rather than blaming
the BRG for its lack of uniqueness, we should use its freedom to
allow the blocks to become correlated in the RG procedure. In our
present approach this correlation will be taken into account in a
``dynamical" way rather than in a ``statistical" way as in the
DMRG. This will be achieved by the introduction of interblock operators
which reflect the ``influence" between neighbour blocks and which
are defined at the boundary of the block in the first step of
our CBRG method.

We have chosen to illustrate our approach the 1D and 2D tight-binding
models mainly for simplicity reasons, but we believe that our
method could be applied to more complicated problems. In fact, the
first step in this direction was already undertaken in reference
\cite{bc-germanyo}, where only 2 states at each stage of the
RG-blocking were retained. This in turn allowed us to obtain the
$1/N^2$ scaling law for the size dependence of the
first-excited-state energy.

 In this paper we shall give the general mathematical structure
underlying the results of reference \cite{bc-germanyo}.
This will allow us to retain more than two states in the RG-truncation
and also to consider the two-dimensional tight-binding model.
In this fashion, we shall recover the $n^2/N^2$ scaling law for the
$n$-th excited state of the 1D model and the scaling law
$\frac{n^2_1+n^2_2}{N^2}$ in the 2D case. These results will then
show that the CBRG method describes correctly the low energy behaviour
of the 1D and 2D Laplacian.

 This paper is organized as follows. In Sect.2 we present the
Correlated Block RG-method and apply it to the 1D Tight-Binding model
with different boundary conditions at the ends of the chain:
Free-Free, Free-Fixed and Fixed-Fixed BC's.
We correctly reproduce in each and every case the corresponding
$n^2/N^2$-scaling
laws ($N \longrightarrow \infty$) for the spectrum of the excited
states.
This is a novel result for it is achieved within the framework of the CBRG
method
in a clear and transparent fashion without having to resort to {\em ad hoc}
mixing of different BC's states in the truncation operator involved in the CBC
method \cite{white-noack}.
In Sect.3 we extend the CBRG method to deal with two-dimensional Hamiltonians
and apply the procedure to successfully solve the 2D tight-binding model
with Free BC's.
Sect.4 is
devoted to conclusions and prospectives.

\vspace{30 pt}

\Section{The CBRG Method: One Dimension}

The problem  we want to study is the one-dimensional Tight-Binding
model in an open chain with different boundary conditions at its ends.
The Hamiltonian for this system takes the following matricial form,

\begin{equation}
 H_{b,b'} = \left( \begin{array}{cccccc} b& -1 & & & & \\
 -1 & 2 & -1& &  & \\
 & -1 &2 & & &  \\ &  &  &  \ddots &  &  \\
 & &  & & 2 & -1  \\
 & &  & & -1 & b'       \label{1} \end{array}              \right)
\end{equation}

\noindent where $b$ and $b'$ take on the values $1$ (or $2$)
corresponding to  Free (or Fixed) BC's respectively. This Hamiltonian
is the discrete version of the Laplacian $H=-\partial^2_x$, while the
Free or Fixed BC's correspond in  the continuum to the vanishing of
the wave function (Fixed BC's) or its spatial  derivative (Free BC's)
at the ends of the chain, i.e.,

\begin{equation}
 \begin{array}{cccc}  b = 2& \Rightarrow & \Psi (0) = 0 & \mbox{Fixed
BC} \\ b = 1& \Rightarrow & \frac{\partial\Psi}{\partial x}(0) = 0
& \mbox{Free BC}
  \label{2} \end{array}              \end{equation}

\noindent and similarly for $b'$ which contains the BC at the other
end of the  chain.

\noindent Hence, altogether there are 4 Hamiltonians of the type in
(\ref{1}), whose  eigenstates and eigenvalues are the subject of our
RG-techniques.

The first step in the RG method is to divide the lattice into blocks
containing  $n_s$ sites each and labeled with and index $p$
($=1,\ldots,N/n_s$). Let us suppose for a moment that we isolate the
$p$th-block from the rest of the lattice so that its dynamics, as an
independent entity, is governed by a Hamiltonian denoted by $A_p$,
which we may call {\em uncorrelated block Hamiltonian}. The
restoration of the block back into the lattice involves two effects.
The first one is that the BC's of the $p$-th block may change  under
the influence of the $p+1$ and $p-1$ blocks. We describe this change
of BC's by the action of  {\em Boundary Operators} denoted by
$B_{p,p\pm 1}$ on the $p$th-block. The second effect is  the
interaction between the $p$th-block and its neighbours $p+1$ and
$p-1$,
 given by interaction  Hamiltonians $C_{p,p\pm1}$ which act
on both $p$ and $p+1$ blocks simultaneously. If the problem  under
consideration is translationally invariant, all the Hamiltonians
defined above are independent of the block label $p$, in which case
we denote them by,

\begin{equation}
 \begin{array}{cc}  A_p = A & \\ B_{p,p+1} = B_R& B_{p,p-1} = B_L \\
C_{p,p+1} = C & C_{p,p-1} = C^{\dagger}
  \label{3} \end{array}              \end{equation}

The $H_{Free, Free}$ Hamiltonian (\ref{1}) gives an example of
this as we shall show below. Hence, for the time being, we shall
consider the situation described by (\ref{3}) and leave the more
general case after explaining the general ideas.

In the standard BRG method the block Hamiltonian $H_B$ and the
interblock  Hamiltonian $H_{BB}$ are given,
according to our  previous definitions,
 by the following formulas

\begin{equation} H_B = A + B_L + B_R  \label{4a}
\end{equation}

\begin{equation}
 H_{BB} =  \left( \begin{array}{cc}  0& C\\ C^{\dagger} &0
  \label{4b} \end{array}      \right)         \end{equation}

\noindent The whole Hamiltonian is by all means the sum of $H_B$ and
$H_{BB}$ for all the blocks of the chain.
For a review on the Block RG method see \cite{jullienlibro} and chapter 11
of reference \cite{jaitisi}.

\noindent Next step in the RG method is to diagonalize $H_B$ and keep
its, say $m$ ($m<n_s$), lowest eigenstates. The truncation is given by a
$n_s \times m$ matrix $T$ whose columns are precisely the components
of the $m$ lowest  eigenstates of $H_B$. The renormalized Hamiltonian
in the new basis is given by,

\begin{equation} H' = T^{\dagger} (H_B + H_{BB}) T  \label{4c}
\end{equation}

\noindent At first sight from Eq. (\ref{4a}) it would seem that we
have taken into account the effect of the BC's on a given block.
However, as the examples show, this is quite a bit illusory. On the
other hand, the distinction among $A$, $B_L$  and $B_R$ is rather
inmaterial as far as $H_B$ is concerned, and in fact  no distinction
of this sort is made in the standard BRG formalism. Finally, let us
observe that $H_B$ and $H_{BB}$ play rather different roles in the
truncation procedure. This asymmetry has been observed as a source of
problems by several authors in the past \cite{fpacheco}, \cite{rabin}.

We shall mention that
 this asymmetry has recently been related to quantum groups in a
fashion  which has led to a new RG method called the Renormalization
Quantum  Group method \cite{q-germanyo}, \cite{qbis-germanyo}.

Therefore, from various points of view, one is urged to make more
explicit the role played by the BC-operators $B_L$ and $B_R$ in our
CBRG procedure. For this purpose, we have found convenient to use the
concept of superblock already introduced in reference
\cite{white-noack}. We shall define a superblock as the set of two
consecutive blocks, $p$ and  $p+1$ and denoted by $(p,p+1)$. The great
advantage of the superblock is that it allows us to  materialize the
distinction among $A$, $B_L$  and $B_R$ . In fact, just as the
isolation of a single block leads us to the definition of the
Hamiltonian $A$, the isolation  of two blocks contained in a
superblock  allows us to define  $B_L$, $B_R$ and also $C$ through
the superblock  Hamiltonian $ H_{sB} $ as follows,

\begin{equation}
 H_{sB} =  \left( \begin{array}{cc}  A + B_R& C\\ C^{\dagger}
 & A + B_L
  \label{5} \end{array}      \right)         \end{equation}

\noindent Similarly, the Hamiltonian describing the interaction
between superblocks is given by (see Fig.1)

\begin{equation}
 H_{sB,sB} =  \left( \begin{array}{cccc}
0& & & \\
& B_R& C&\\
&C^{\dagger} &  B_L& \\
& & &0
  \label{6} \end{array}      \right)         \end{equation}

\noindent Now instead of diagonalizing $H_B$ in Eq. (\ref{4a}), in the
CBRG method we shall diagonalize $H_{sB}$ in Eq. (\ref{5}), and
afterwards keep
 the $m=n_s$  lowest eigenstates in the tight-binding  model.
As in the standard BRG method,  the change to the truncated basis
defines the renormalized operators as follows:

\begin{equation} H_{sB} \longrightarrow T^{\dagger} H_{sB} T = A'
\label{7a}             \end{equation}

\begin{equation} H_{sB,sB} \longrightarrow T^{\dagger}
H_{sB,sB} T=  \left(
\begin{array}{cc}  B'_R& C'\\ \mbox{$C'$}^{\dagger} &B'_L
  \label{7b} \end{array}      \right)         \end{equation}

\noindent where the matrices $A'$,$B'_R$, $B'_L$ and $C'$ are the
renormalized version of the operators $A$,$B_R$, $B_L$ and $C$, and
they exhibit the same  geometrical interpretation for the
renormalized block as their unprimed partners for the original
blocks.

If we set $B_R = B_L = 0$ in Eqs. (\ref{5}) and (\ref{6}), then after
the first RG-step we get $B'_R = B'_L = 0$ and thus the previous
RG-scheme coincides with the  standard BRG. We may say that
uncorrelated blocks are in a sense  a fixed point of our method.
However, this fixed point may be unstable, and to explore this
possibility
one has to look for non-vanishing $B$-operators and their
RG-evolution.

Let us address now some examples. We shall first study the
Hamiltonian (\ref{1}) with Free BC's at the ends ($b=b'=1$).
Choosing $n_s=3$ for example, we see that the choice for the
operators $A$,$B_R$, $B_L$ and $C$ in the first step of the CBRG
procedure is given by,

\begin{equation} A =  \left( \begin{array}{ccc}
 1& -1 &0 \\ -1 & 2
& -1\\
 0 & -1 & 1
   \end{array}      \right)  , \  B_R =  \left( \begin{array}{ccc}
0&  & \\
 & 0 & \\
  &  & 1
   \end{array}      \right)  , \  B_L =  \left( \begin{array}{ccc}
1&  & \\
 & 0 & \\
  &  & 0
   \end{array}      \right)  , \  C =  \left( \begin{array}{ccc}
0& 0 &0 \\ 0 & 0 & 0\\
 -1 & 0 & 0
   \end{array}      \right)   \label{8}
      \end{equation}

\noindent This choice is equivalent to the assumption that an
isolated block satisfies Free BC's at its ends. The role of $B_R$
and $B_L$ is to {\em join} these blocks into a single chain. This is
the  {\em geometrical} explanation of Eqs. (\ref{8}). In more
general cases one must have to explore which is the best  choice.
The generalization of Eqs.(\ref{8}) to blocks with more  than 3
sites is obvious. In Table 1 we collect our CBRG-results for the first 5
excited states for a chain of $N=12 \times 2^6=768$ sites.
Comparison with exact results gives a good agreement.

 An important feature of our CBRG method is that the
$n^2/N^2$- scaling
law ($N
\longrightarrow \infty$) for the energy of the $n$-excited states of
a chain made up of $N$ sites, is reproduced correctly (see Fig.2).
In Table 2 we show the variation of the first-excited-state energy
with the size $N$ of the chain. From those values we can extract the
corresponding $1/N^2$-law which turns out to be,

\begin{equation}
 E^{(CBRG)}_1 (N) = c^{(1)}_{CBRG} \frac{1}{N^2}, \ \ c^{(1)}_{CBRG} = 9.8080,
 \ (N \longrightarrow \infty) \ \mbox{Free-Free BC's}\label{8a}
  \end{equation}

\noindent while the exact value for the proportionality constant $c$
is $c_{\mbox{exact}}=\pi^2=9.86$. This amounts to a 0.6 \% error.

\noindent Likewise, we have  enough data so as to obtain the
corresponding $n^2/N^2$-law for the whole set of 5 excited states.
Thus, the scaling law we obtain is,

\begin{equation}
 E^{(CBRG)}_n (N) = c_{CBRG} \frac{n^2}{N^2}, \ \ c_{CBRG} = 8.4733,
 \ (N \longrightarrow \infty) \ \mbox{Free-Free BC's}\label{8ab}
  \end{equation}

\noindent which now amounts to a 7.34 \% error.
This is a natural fact from the worse knowledge of the highest excited states
 of the spectrum in a RG-scheme.

We can make even more explicit the successful achievement of the
$1/N^2$-scaling law by leaving as a free adjustable parameter the
exponent of $1/N$ in addition to the proportionality constant.
Let us denote by $\theta $ this critical exponent. Using data from
20 to 50 steps of our CBRG-method for several truncation of states
according to our scheme $2 n_s \rightarrow n_s$
(namely, $n_s=10,13,20$)
we arrive at the following results,

\begin{equation}
 E^{(CBRG)}_1 (N) = c_{CBRG} \frac{1}{N^{\theta}}, \ \
 \ (N \longrightarrow \infty) \ \mbox{Free-Free BC's}\label{8abc}
  \end{equation}

\begin{equation}
 \begin{array}{ccc}
\mbox{For}\ \  20 \longrightarrow 10&  &\theta = 1.9708 \\
 \mbox{For}\ \  26 \longrightarrow 13&  &\theta = 1.9734 \\
 \mbox{For}\ \ 40 \longrightarrow 20&  &\theta = 1.9854
   \end{array}
 \label{8d}
      \end{equation}

\noindent These results clearly support the fact that we have
correctly reproduced the exact value of $\theta = 2$ for the
finite-size critical exponent.

\noindent Last, but not least, as was proved in \cite{bc-germanyo} our
CBRG method gives the {\em exact} energy of the ground state for every
step of the RG-procedure for Free-Free BC's.

\noindent In reference \cite{bc-germanyo} it was shown that one can
reproduce easily the wave function of the excited states. This
procedure was called {\em reconstruction} since it works ``downwards"
in the CBRG method. The basic equation to be used is the  {\em
reconstruction equation} \cite{bc-germanyo},

\begin{equation} \Psi^{(r+1)} = L_r \Psi^{(r)}_L + R_r
\Psi^{(r)}_R\label{9}
  \end{equation}

\noindent where $\Psi^{(r)}$ denotes the collection of $m$ lowest
eigenstates in the $r$-step of the CBRG-procedure, and $L_r$, $R_r$
are the block matrices in terms of which the truncation matrix
$T^{\dagger}$
can be written as $T^{\dagger} = (L_r,R_r)$.

\noindent Our results for a chain of $N=12 \times 2^6=768$
sites and $n_s=6$ states kept are given in
Fig.3 where we have plotted the first 5 excited states and compare them
with the exact wave functions. There are some remarkable facts regarding
these figures.
Firstly,  the number of nodes is correctly preserved by our CBRG wave
functions.
Secondly, the Free-Free type of boundary conditions are also correctly
reproduced
at the ends of the chain.
And lastly, it is worthwhile to point out that the CBRG wave functions
``degrade gracefully" as the energy of the excited state raises in
accordance with the fact that the lower the energy is, the more
reliable are the results.

This ends the results for the Free-Free BC's. In order to address
other types of BC's we must come back to the case where the
matrices $A$,$B_R$, $B_L$ and $C$ depend on each particular block.
Thus, for example, for the Fixed-Free BC's we shall choose as  the
uncorrelated $A$-matrix for the block located to the left end  of
the chain the following form ($n_s=3$),

\begin{equation} A_1 =  \left( \begin{array}{ccc}   2& -1 &0 \\ -1 &
2 & -1\\
 0 & -1 & 1
   \end{array}      \right) \ \ \mbox{Fixed-Free BC's} \label{10}
      \end{equation}

\noindent while the remaining matrices $A_p$, $(p=2,\ldots,N/3)$,
will be given by Eqs.(\ref{3}), (\ref{8}).

For Free-Fixed BC's, it is the last $A$-matrix which we have to take
different from the others, namely,

\begin{equation} A_{N/3} =  \left( \begin{array}{ccc}   1& -1 &0 \\
-1 & 2 & -1\\
 0 & -1 & 2
   \end{array}      \right) \ \ \mbox{Free-Fixed BC's} \label{11}
      \end{equation}

\noindent As for the Fixed-Fixed BC's case, we must change the $A$-matrix
at both ends of the chain according to the following prescription,

\begin{equation} A_1 =  \left( \begin{array}{ccc}   2& -1 &0 \\ -1 &
2 & -1\\
 0 & -1 & 1
   \end{array}      \right), \ \ A_{N/3} =  \left( \begin{array}{ccc}   1& -1
&0 \\
-1 & 2 & -1\\
 0 & -1 & 2
   \end{array}      \right) \ \
 \mbox{Fixed-Fixed BC's} \label{11a}
      \end{equation}

\noindent Then we follow the same steps as for the Free-Free BC's,
taking care that the  $A$,$B_R$, $B_L$ and $C$ matrices in each
CBRG-step may depend on  the position of the blocks. This implies
in particular that the embedding $T$-matrices may also vary from
block to block.

 In Tables 3 and 4 we  summarize our results for the Free-Fixed
and Fixed-Fixed BC's (Fixed-Free BC's are equivalent to Free-Fixed
BC's by parity transformation).
In these tables we present our CBRG results for the first 6 lowest lying states
for the 1D tight-binding model in a chain of $N=12 \times 2^5=384$ sites
with mixed boundary conditions, and they are compared against the exact
and standard BRG values.
Several remarks are in order.
First, we observe that the CBRG method produces  a good agreement with the
exact results and certainly much more accurate by several orders of
magnitude than the old BRG method.
Second, the CBRG method is able to reproduce the corresponding
$n^2/N^2$-scaling
laws for the spectrum of excited states in each case of mixed BC's.
Namely,

\begin{itemize}

\item For Free-Fixed BC's and considering just the ground state, we have

\begin{equation}
 E^{(CBRG)}_0 (N) = c^{(0)}_{CBRG} \frac{1}{4N^2}, \ \
c^{(0)}_{CBRG} = 9.072,
 \ (N \longrightarrow \infty) \ \mbox{Free-Fixed BC's}\label{11b}
  \end{equation}

\noindent which amounts to a 8 \% error with respect to the exact value of
$c_{exact}=\pi^2$.

\noindent As for the corresponding law for the whole spectrum, we find

\begin{equation}
 E^{(CBRG)}_n (N) = c_{CBRG} \frac{(n+1)^2}{4N^2}, \ \
c_{CBRG} = 7.6729,
 \ (N \longrightarrow \infty) \ \mbox{Free-Fixed BC's}\label{11c}
  \end{equation}

\noindent which represents a 11.5 \% error with respect to the exact value of
$\pi^2$.

\item For Free-Fixed BC's and considering just the ground state,
 we have

\begin{equation}
 E^{(CBRG)}_0 (N) = c^{(0)}_{CBRG} \frac{1}{N^2}, \ \
c^{(0)}_{CBRG} = 8.35,
 \ (N \longrightarrow \infty) \ \mbox{Fixed-Fixed BC's}\label{11ddd}
  \end{equation}

\noindent which amounts to a 8 \% error with respect to the exact value of
$c_{exact}=\pi^2$.

\noindent As for the corresponding law for the whole spectrum, we find

\begin{equation}
 E^{(CBRG)}_n (N) = c_{CBRG} \frac{(n+1)^2}{N^2}, \ \
c_{CBRG} = 6.9696,
 \ (N \longrightarrow \infty) \ \mbox{Fixed-Fixed BC's}\label{11dd}
  \end{equation}

\noindent which represents a 16 \% error with respect to the exact value of
$\pi^2$.

\end{itemize}

\noindent We obtain bigger errors in the determination of these
scaling laws as compared with the Free-Free case mainly because
we have used less data in our fitting. Nevertherless, we find a
good agreement with the exact results. Yet, there is another reason
as to why the accuracy in the case of mixed BC's is worse, namely,
the ground state wave function $\Psi_0$ is not homogeneous in space
as it is in the Free-Free case \cite{bc-germanyo}.
This makes the RG-procedure more involved and a source of extra
uncertainties.

\noindent Let us mention in passing that we are also able to
make a wave function
reconstruction in the mixed BC's cases as has been done for
the Free-Free BC
case.

The outcome of all the results presented so far is that we
have succeded in
devising a Real-Space RG method capable of reproducing the
correct eigenvalues
and eigenstates for the tight-binding model as originally
envisaged by Wilson, within a certain accuracy which can in
principle be improved.

Althoug the model we have employed to test our CBRG-method is a
tight-binding model, there are some remarkable facts regarding
the fixed-point structure of our CBRG-solution that we would like
to stress.
Namely, we have found that after enough number of CBRG-iterations,
the matrices $A$, $B_L$, $B_R$ and $C$ in the Free-Free case scale
nicely with the size $N$ of the chain according to the dynamical
critical exponent $z$. To be more precise, let us introduce
the fixed point values of those matrices denoted by
$A^{\ast}$, $B^{\ast}_L$, $B^{\ast}_R$ and $C^{\ast}$ which we
define as,

\begin{equation}
A^{\ast} = N^{-z} a^{\ast}, \ \
B^{\ast}_L = N^{-z} b^{\ast}_L, \ \
B^{\ast}_R = N^{-z} b^{\ast}_R, \ \
C^{\ast} = N^{-z} c^{\ast},  \ \ \mbox{Fixed-Point values}\label{11f}
\end{equation}

\noindent in terms of the scaled matrices
$a^{\ast}$, $b^{\ast}_L$, $b^{\ast}_R$ and $c^{\ast}$. For a block
of 3 sites ($n_s=3$) we find the following Fixed-Point structure
parametrized by two constants $s$ and $t$ (for bigger $n_s$ we
need extra parameters),

\begin{equation} a^{\ast} =  0, \
 b^{\ast}_R =  \left( \begin{array}{ccc}
1& s & s\\
 s& t &t \\
  s& t & t   \end{array}      \right)  , \
b^{\ast}_L =  \left( \begin{array}{ccc}
1&  -s& s\\
-s & t & -t\\
  s& -t & t
   \end{array}      \right)  , \
c^{\ast} =  \left( \begin{array}{ccc}
-1& s &-s \\
-s & t & -t\\
 -s & t & -t
   \end{array}      \right)   \label{11ff}
      \end{equation}

\noindent with $s=1.3993$ and $t=1.9581$. The critical exponent $z$
we obtain is,

\begin{equation}
z = 0.9999 \label{11fff}
      \end{equation}

\noindent which is indeed very close to the exact value $z=1$
(actually, it differs in the ninth decimal digit).

\noindent The interpretation of this Fixed-Point in the context of
the CBRG method is as follows. We pointed out before that when the
boundary operators $b_{L,R}$ vanish we recover the standard BRG
method in which the blocks are not correlated. Here we find that it
is the uncorrelated Hamiltonian which vanish, while the boundary
$b_{L,R}$ and interaction $c$ operators do not vanish within the
scaling law.
This fact may perhaps be interpreted by saying that in the example
under study the correlation between blocks is more important
than their selfenergy. In references \cite{nishino},
\cite{ostlund-rommer}, \cite{nishino-okunishi} it was shown that the
DMRG method leads, in the thermodynamic limit, to a
``product form" ansatz for the
ground state wave function. In our case we see from Eqs.(\ref{11f}),
(\ref{11ff}) that we also reach thermodynamical limit, which leads us
to ask about the nature of the ansatz for the ground state and
excited states implied by the CBRG method. The answer to this
question will be addressed in a future publication but it suffices
to say that both the DMRG and the CBRG methods seem to yield
different ansatzs of the ground state wave function.
In a few words, the DMRG is associated with a ``vertex picture"
while the CBRG is associated with a ``string picture".

\Section{The Two-Dimensional CBRG-Algorithm}

The RG-method that we have devised in the one-dimensional problem
can be generalized in a natural way to higher dimensions. We shall
consider for simplicity the 2D case. First of all, we divide the
square lattice into blocks of $n_s$  sites each. Each block will in
turn be a square lattice with  a minimum of 4 sites ($= 2\times 2$
block). As in 1D, we shall define the following Hamiltonians to
carry out the CBRG-program,

\begin{itemize}

\item $A_p =$ self-energy of the $p$-th block isolated from the
lattice.

\item $B_{p,q} =$ self-energy of the $p$-th block induced by the
presence of the $q$-th block.

\item $C_{p,q} =$ interaction between the $p$-th block and the
$q$-th block.

\end{itemize}

\noindent The difference with respect to the 1D case is that each
block has now 4 neighbours  and therefore there are four different
$B$ and $C$ matrices.

Let us consider again the Hamiltonian of a free particle moving  in
a 2D-box with Free BC's at the boundaries of the box. The 2D
Hamiltonian is given again by the incidence matrix of the  lattice.
As in 1D we shall choose the matrix $A$ as the incidence matrix of
the block. Thus, for example, for a $2\times 2$ block labelled as in
Fig.4 we have,

\begin{equation} A =  \left( \begin{array}{cccc}   2 & -1 & 0 & -1\\
-1 & 2 & -1 & 0\\ 0 & -1 & 2 & -1\\ -1 & 0 & -1 & 2
   \end{array}      \right) \label{12}
      \end{equation}

\noindent The 4 Boundary Operators $B$ in the same basis as
in Fig.5 a
have a  diagramatic representation as shown in Fig.4 which helps us
to keep track of their location in the block $H_B$ and interblock
$H_{BB}$ Hamiltonians. Dotted line in Fig.4 means {\em influence} or
{\em correlation} between neighbouring blocks. Their explicit
matricial form is as follows,

\begin{equation} B_{12} = B_{43} = B_L =  \left(
\begin{array}{cccc}   0 &  &  & \\
 & 1 &  & \\
 &  & 1 & \\
 &  &  & 0
   \end{array}  \right), \ \  B_{21} = B_{34} = B_R =  \left(
\begin{array}{cccc}   1 &  &  & \\
 & 0 &  & \\
 &  & 0 & \\
 &  &  & 1
   \end{array}  \right) \label{13}
      \end{equation}

\begin{equation} B_{14} = B_{23} = B_D =  \left(
\begin{array}{cccc}   0 &  &  & \\
 & 0 &  & \\
 &  & 1 & \\
 &  &  & 1
   \end{array}  \right), \ \  B_{41} = B_{32} = B_U =  \left(
\begin{array}{cccc}   1 &  &  & \\
 & 1 &  & \\
 &  & 0 & \\
 &  &  & 0
   \end{array}  \right) \label{14}
      \end{equation}

\noindent where the labels denote the position of the neighbouring
blocks and we have used the translation invariance of the 2D
tight-binding model so that we need only to distinguish between
Right and Left, and Up and Down.

\noindent As for the Interaction $C$-Operators whose diagramatic
representation is depicted in Fig.4 we have the following matricial
representation, with the same considerations as for the
$B$-operators,

\begin{equation} C_{12} = C_{43} = C_{LR} =  \left(
\begin{array}{cccc}   0 & 0 & 0 & 0\\ -1 & 0 & 0 & 0\\
 0& 0 & 0 & -1\\
 0&  0& 0 & 0
   \end{array}  \right), \ \  C_{21} = C_{34} = C_{RL} =  \left(
\begin{array}{cccc}   0 & -1 & 0 & 0\\ 0 & 0 & 0 & 0\\
 0& 0 & 0 & 0\\
 0&  0& -1 & 0
   \end{array}  \right) \label{15}
      \end{equation}

\begin{equation} C_{14} = C_{23} = C_{DU} =  \left(
\begin{array}{cccc}   0 & 0 & 0 & 0\\ 0 & 0 & 0 & 0\\
 0& -1 & 0 & 0\\
 -1&  0& 0 & 0
   \end{array}  \right), \ \  C_{41} = C_{32} = C_{UD} =  \left(
\begin{array}{cccc}   0 & 0 & 0 & -1\\ 0 & 0 & -1 & 0\\
 0& 0 & 0 & 0\\
 0&  0& 0 & 0
   \end{array}  \right) \label{16}
      \end{equation}

\noindent Thus translation invariance reduces the number of
independent CBRG-matrices by a half. These relations are particular
of the problem at hand but we must left open the posibility of
having all those matrices different from each other in order to
handle more complicated problems.

\noindent Now that we have all the elements entering in our
CBRG-method we proceed to construct the block $H_{sB}$ and interblock
$H_{sB,sB}$
 Hamiltonians out of them. To this end we have to consider a
superblock made up of 4 blocks as shown in Fig.5 a) where the basis
chosen is made explicitly. Thus, for $H_{sB}$ we have,

\begin{equation} H_{sB} =  \left( \begin{array}{cccc}   A + B_L +
B_D & C_{LR} & 0 & C_{DU}\\
C_{RL}  & A + B_R + B_D & C_{DU}& 0\\
 0& C_{UD} & A + B_R + B_U & C_{RL}\\
 C_{UD} &  0& C_{LR} & A + B_L + B_U
   \end{array}  \right) \label{17}
      \end{equation}

\noindent This is a $4 n_s \times 4 n_s$ matrix made up of $ n_s
\times  n_s$  matrices.

\noindent As for the interblock Hamiltonian $H_{sB,sB}$ we have to
distinguish between $(sB,sB)$-couplings of horizontal type denoted by
$H^{(hor)}_{sB,sB}$ and vertical type denoted by
$H^{(ver)}_{sB,sB}$, which read explicitly as, \begin{equation}
H^{(hor)}_{sB,sB} =  \left( \begin{array}{cccccccc}   B_R &  &  &  &
0 & C_{RL} & 0 & 0\\
 & 0 & &  & 0 & 0 & 0 & 0\\
 & & 0 &  & 0 & 0 & 0 & 0\\
 & & & B_R & 0 & 0 & C_{RL} & 0 \\ 0 & 0 & 0 & 0 & 0 &  &  & \\
 C_{LR}& 0 &0 & 0 &  & B_L& & \\
 0& 0& 0 & C_{LR} &  &  & B_L & \\
 0& 0& 0& 0 &  &  &  & 0
   \end{array}  \right) \label{18a}
      \end{equation}

\begin{equation} H^{(ver)}_{sB,sB} =  \left(
\begin{array}{cccccccc}   B_D &  &  &  & 0 & 0 & 0 & C_{UD}\\
 & B_D & &  & 0 & 0 & C_{UD} & 0\\
 & & 0 &  & 0 & 0 & 0 & 0\\
 & & & 0 & 0 & 0 & 0 & 0 \\ 0 & 0 & 0 & 0 & 0 &  &  & \\
 0& 0 &0 & 0 &  & 0& & \\
 0& C_{DU}& 0 & 0 &  &  & B_U & \\
 C_{DU}& 0& 0& 0 &  &  &  & B_U
   \end{array}  \right) \label{18b}
      \end{equation}

\noindent where we have made use again of translational invariance.

\noindent Once that we have made our choice for the decomposition of
the total Hamiltonian of the 2D-tight-binding model into block and
interblock Hamiltonians according to our CBRG-prescription, we can
carry on with the truncation part of the RG-method. We shall keep
$n_s$ states out of $4 n_s$ states per superblock so that our
truncation scheme may be summarized as,

\[ 4 n_s \ \mbox{(superblock)} \longrightarrow n_s \ \mbox{(new
block)} \]

Recall that at each step of the CBRG-method we need to identify the
$A$, $B_L$, $B_R$ and $C$ operators which define the truncation
procedure for  the next step of the method. For this purpose,
firstly the truncation of the  superblock $H_{sB}$ gives rise to the
$A'$ uncorrelated self-energy operator for the next RG-step, namely,

\begin{equation} H_{sB}  \  \mbox{( $4 n_s \times 4 n_s$
matrix)}\longrightarrow A' \  \mbox{($n_s \times n_s$ matrix)}
\label{19}
      \end{equation}

\noindent To identify the rest of the operators we have to
renormalize the interblock Hamiltonian which comes in two types,
horizontal and vertical. As exemplified schematically in Fig.6 the
renormalization of the $H^{(hor)}_{sB,sB}$ Hamiltonian is given by,

\begin{equation} H^{(hor)}_{sB,sB} \longrightarrow  \left( \begin{array}{cc}
B'_R & C'_{RL}  \\
 C'_{LR}& B'_L
   \end{array}  \right) \label{20}
      \end{equation}

\noindent Likewise, for the $H^{(ver)}_{sB,sB}$ Hamiltonian we have
(see Fig.6),

\begin{equation} H^{(ver)}_{sB,sB} \longrightarrow  \left( \begin{array}{cc}
B'_D & C'_{UD}  \\
 C'_{DU}& B'_U
   \end{array}  \right) \label{21}
      \end{equation}

\noindent Now that we have identified all the operators defining the
CBRG method at the new stage of the renormalization, we may
reconstruct the new superblock Hamiltonian $H'_{sB}$, which in turn
has the same form as the original  $H_{sB}$ in Eq.(\ref{17})
substituting all the operators by their {\em primed versions}.
This statement can be explicitly checked by considering the
set of 4 superblocks as depicted in Fig.5.
Firstly, the new $H'_{sB}$ has a contribution coming from the
truncation of each of the 4 superblocks in Fig.5, each of them
contributing with an $A'$-operator as in Eq.(\ref{19}).
Secondly, $H'_{sB}$ picks up two more contributions coming from
the horizontal and vertical interaction between superblocks,
which we denote by $H_{\leftrightarrow}$ and $H_{\updownarrow}$.
Thus, in the CBRG-method $H'_{sB}$ is renormalized as,

\[
H'_{sB} =  \left( \begin{array}{cccc}
A'  &  &  & \\
  & A'  & & \\
 & & A' & \\
  &  &  & A'
   \end{array}  \right) \ \ \longleftarrow
\mbox{(single superblock contribution)}
\]
\[
(H_{\leftrightarrow}) \ \rightarrow \ +
\left( \begin{array}{cccc}
B'_L  & C'_{LR} &  & \\
C'_{RL}  & B'_R  & & \\
 & & 0 & \\
  &  &  & 0
   \end{array}  \right) +
\left( \begin{array}{cccc}
0  &  &  & \\
  & 0  & & \\
 & & B'_R &C'_{RL} \\
  &  & C'_{LR} & B'_L
   \end{array}  \right)
\]
\begin{equation}
(H_{\updownarrow}) \ \rightarrow \ +
\left( \begin{array}{cccc}
B'_U  &  &  & C'_{DU}\\
  &  0 & & \\
 & & 0 & \\
 C'_{UD} &  &  & B'_D
   \end{array}  \right) +
\left( \begin{array}{cccc}
0  &  &  & \\
  & B'_U  & C'_{DU} & \\
  & C'_{UD} &B'_D & \\
  &  &  & 0
   \end{array}  \right)
\label{23}
      \end{equation}

\noindent and altogether we arrive at the previously stated result
of Eq.(\ref{17}).

\noindent Similarly we may proceed with the renormalized
interblock Hamiltonians $\mbox{$H'$}^{(hor)}_{sB,sB}$ (\ref{20}) and
$\mbox{$H'$}^{(ver)}_{sB,sB}$ (\ref{21})
and we end up with the same form for them as the original ones.

\noindent This ends the implementation of the CBRG-method for the
2D-tight-binding model.

In Table 5 we collect our CBRG results for the first 4 lowest lying
states for a chain of $N=4\times 4\times 4^6=65536$ sites. Comparison
with the exact results gives a good agreement. We have also data
from truncations with blocks of $n_s=9$ and $n_s=16$ sites which
enforce this statement.
Moreover, notice that the first excited state is a doublet as in the
exact solution.

Another important result of our CBRG-method is that the
$(n^2_1 + n^2_2)/N^2$ scaling law for the energy of the
$(n_1,n_2)$-excited states of a square lattice of length $N$ is
reproduced correctly. In fact, from data of the $n_s=16$ sites
truncation for the first-excited-state energy we can extract the
corresponding $1/N^2$-scaling law which turns out to be,

\begin{equation}
 E^{(CBRG)}_1 (N) = c^{(1)}_{CBRG} \frac{1}{N^2}, \ \
c^{(1)}_{CBRG} = 9.7365,
 \ (N \longrightarrow \infty) \ \mbox{D=2 Free BC's}\label{24}
  \end{equation}

\noindent while the exact value of the proportionality constant
$c$ is $c_{exact}=\pi^2=9.86$. This amounts to a 1.3 \% error.

Likewise, we may obtain the full $(n^2_1 + n^2_2)/N^2$ scaling law
for the whole set of 15 excited states and we find,

\begin{equation}
 E^{(CBRG)}_{(n_1,n_2)} (N) = c^{(1)}_{CBRG}
\frac{(n^2_1 + n^2_2)}{N^2}, \ \
c_{CBRG} = 7.9074,
 \ (N \longrightarrow \infty) \ \mbox{D=2 Free BC's}\label{25}
  \end{equation}

\noindent which now amounts to a 10.5 \% error.

As in the 1D Free-Free case, we can determine critical scaling
exponent $\theta$ (\ref{8abc}). For a truncation scheme
$16 \rightarrow 4$ we find,

\begin{equation}
 \theta = 1.99999981 \ \ \ \mbox{D=2}\label{26}
  \end{equation}

\noindent which clearly supports the scaling laws introduced above.
Notice again (see Table 5) that our CBRG method gives the exact
(within machine precision) energy of the ground state. This is true
for every step of the RG, as was proved in \cite{bc-germanyo} for 1D.

We can also perform the wave function reconstruction of the excited
states in the two-dimensional real space. This is achieved by a
two-dimensional extension of the reconstruction equation (\ref{9}).
As an illustration of how the CBRG method performs with this matter,
in Fig.7 we show a two-dimensional density plot of the wave function
reconstruction for the siglet (1,1)-excited state in a square lattice
of $N=65536$ sites with Free BC's. In that figure we show level
lines separating regions of different height which are colored
differently (darker grey levels meaning higher height). This density
plot is made by projecting the 3D wave function representation onto
the $x-y$ plane according to the height of the function. Thus, we
may appreciate two maxima (black spots) and two minima (white spots)
at the corners of the lattice, as it should be according to the exact
solution.

\noindent Recall that the exact solution for the (1,1)  state has
two lines of nodes, one vertical and one horizontal, which pass
through the mid-points of each lattice side. These two node-lines
are the analogue of the node-points of the 1D excited states in
Fig.3. We have also obtained these node-lines correctly within
our method.
Thus, the qualitative real-space form of the excited-state wave
functions are captured by the CBRG procedure.

\Section{Conclusions and Prospectives}

We have presented in this paper a new Real-Space Renormalization
Group method called Correlated Block RG (CBRG) which has been
very much inspired by the Density Matrix RG method introduced
by White \cite{white}, \cite{white-noack}. The basic notion developed
by White in his novel treatment of Real-Spaces RG methods is that
blocks are not isolated during the process of truncation of states
and that they are certainly correlated. We have made a novel
treatment of this concept without having to resort to a density
matrix formalism nor to Wilson's onion-scheme for enlarging
the lattice (chain).

The way in which we take into account correlation between blocks
is a dynamical one. This means that we start the first step of
our CBRG method defining uncorrelated block Hamiltonians $A$,
boundary operators $B$ and interaction operators $C$ which are used
to construct the block Hamiltonian $H_B$ and thereby the
truncation operator $T^{\dagger}$. This construction is reproduced
at each step of the RG procedure. Thus, it is the system under
consideration which goes choosing  the correlation
effects between blocks during the renormalization process.

 We have tested the CBRG method on the 1D and 2D Tight-Binding
model with different boundary conditions and obtained very good
results for the spectrum of low lying states and its real space
representation of their wave functions. Moreover, we have correctly
reproduced the scaling laws $n^2/N^2$ for the energy of the excited
states in 1D for Free-Free, Free-Fixed and Fixed-Fixed BC's
and the scaling law $(n^2_1 + n^2_2)/N^2$ for Free BC's in 2D.
In addition, the correct dynamical exponent $z$ is also reproduced.
Therefore, we believe that we have definitely solved in a transparent
and systematic way the boundary problem for Tight-Binding models posed
by Wilson \cite{wilsontalk} in his search for a better understanding
of how a Real-Space RG method works.

  In order to have a more extensive comprenhension of the correlations
introduced in our CBRG method, we think it would be enlightning to
make a thorough comparison between the process of truncation of
states in our method and the one employed in the DMRG method by
means of the onion-scheme. This issue is left for future work.
Another important line of study is the extension of the present
CBRG method to interacting systems appearing in many-body problems.
In this regard, we are currently studying a many-body system of
fermions in one dimension as another prospective use of our method.


\vspace{20 pt}

{\bf Acknowledgements}

One of us (G.S.) would like to thank T. Nishino for useful discussions.
 \vspace{20 pt}

Work partially supported in part by CICYT under  contracts AEN93-0776
(M.A.M.-D.) and the Swiss National Science Foundation and by the
Spanish Fund DGICYT, Ref. PR95-284
(G.S.).


\def\baselinestretch{1.5} \noindent  %

\newpage \section*{Table captions}

{\bf Table 1 :} Exact and CBRG Values of  Low Lying
States for the 1D Tight-Binding  Model for a chain of
$N=12 \times 2^6=768$ sites with Free-Free BC's.

{\bf Table 2 :} Exact  and new CBRG values
of the first excited state
 for the 1D Tight-Binding  Model with Free-Free BC's.

{\bf Table 3 :} Exact, Standard RG and CBRG Values of  Low Lying
States for the 1D Tight-Binding  Model for a chain of
$N=12 \times 2^5=384$ sites with Free-Fixed BC's.

{\bf Table 4 :} Exact, Standard RG and CBRG Values of  Low Lying
States for the 1D Tight-Binding  Model for a chain of
$N=12 \times 2^5=384$ sites with Fixed-Fixed BC's.

{\bf Table 5 :} Exact and CBRG Values of  Low Lying
States for the 2D Tight-Binding  Model for a lattice of
$N=4\times 4 \times 4^6=65536$ sites with Free BC's.

\newpage \section*{Figure captions} \noindent

 {\bf Figure 1:} Pictorical decomposition of a given Hamiltonian
$H$ into
uncorrelated  $A$-operators, correlation $B_L$- $B_R$-operators
and interaction $C$-operators according to the CBRG method. $B_1$
is a superblock made up of two $L_1$ and $R_1$ blocks.

 {\bf Figure 2 :} The $n^2/N^2$-law
for the first
5 excited states of the 1D Tight-Binding  Model for a chain of
$N=12 \times 2^m$ sites with Free-Free BC's. This is a
$\ln E_n$-$\ln m$ plot.

 {\bf Figure 3 :} The  wave function reconstruction for the first
5 excited states of the 1D Tight-Binding  Model for a chain of
$N=12 \times 2^6=768$ sites with Free-Free BC's. We have scaled
up the exact results by a factor of 1.23 for clarity.

 {\bf Figure 4 :} Pictorical representation of
uncorrelated  $A$-operators, correlation $B_L$- $B_R$-
$B_U$ $B_D$-operators
and interaction $C_{LR}$-
$C_{UD}$-operators according to the CBRG method
in D=2 dimensions.

{\bf Figure 5  :} a) Two-dimensional superblock decomposition
into 4 blocks of $2\times 2$ sites showing the choice of basis.
b) Vertical and horizontal interactions between 4 neighbouring
superblocks in D=2 dimensions.

{\bf Figure 6 :} Vertical and horizontal renormalizations of
neighbouring
superblocks in D=2 dimensions.

 {\bf Figure 7 :} Plot of the wave
function reconstruction for the  singlet (1,1)-excited state
in the 2D Tight-Binding  Model for a lattice of
$N=4\times 4 \times 4^6=65536$ sites with Free BC's.

\newpage



\begin{table}[p]
\centering
\begin{tabular}{|c|c|c|}
\hline \hline
  \mbox{Energies} & \mbox{ Exact}
&  \mbox{CBRG} \\  \hline \hline
 $E_0$& $0 $  &
$1.1340\times 10^{-14}$  \\ \hline
$E_1$ & $1.6733 \times 10^{-5}$  & $1.9752\times
10^{-5}$  \\  \hline
 $E_2$ & $6.6932 \times 10^{-5}$  &
$7.6552\times 10^{-5}$ \\ \hline
 $E_3$& $1.5060 \times 10^{-4}$  &
$1.8041\times 10^{-5}$  \\ \hline
 $E_4$& $2.6772 \times 10^{-4}$  &
$2.9681\times 10^{-4}$  \\ \hline
 $E_5$& $4.1831 \times 10^{-4}$   &
$5.1078\times 10^{-4}$   \\ \hline \hline
\end{tabular}
\caption{Exact and CBRG Values of  Low Lying
States for the 1D Tight-Binding  Model for a chain of
$N=12 \times 2^6=768$ sites with Free-Free BC's.}  \end{table}

\begin{table}[p]
\centering
\begin{tabular}{|c|c|c|c|}
\hline \hline
  \mbox{m} & \mbox{N=12 $2^m$}    & $E_1^{(exact)}(N)$ &
$E_1^{(CBRG)}(N)$ \\  \hline \hline
 $0$& $12$ & $6.8148\times 10^{-2}$& $6.8148\times 10^{-2}$\\ \hline
$1$ & $24$ & $1.7110 \times 10^{-2}$
&$1.7375\times 10^{-2}$  \\  \hline
 $2$ & $48$& $4.2826 \times 10^{-3}$  &
$4.4694\times 10^{-3}$  \\ \hline
$3$ &$96$ & $1.0708\times 10^{-3} $ &
$1.1515\times 10^{-3}$  \\  \hline
$4$ & $192$ & $2.6772 \times 10^{-4}$ & $2.9681 \times
10^{-4}$  \\  \hline
$5$ & $384$ & $6.6932 \times 10^{-5}$
& $7.6552 \times 10^{-5}$    \\  \hline
$6$ & $768$ & $1.6733 \times 10^{-5}$ &
 $1.9752 \times 10^{-5}$    \\  \hline
  & $\gg 1$ & $\pi^2/N^2$ & $9.8080/N^2$ \\ \hline \hline
\end{tabular} \caption{Exact  and new CBRG values
of the first excited state
 for the 1D Tight-Binding  Model with Free-Free BC's.}  \end{table}



\begin{table}[p]
\centering
\begin{tabular}{|c|c|c|c|}
\hline \hline
  \mbox{Energies} & \mbox{ Exact}    & \mbox{ Standard BRG}
&  \mbox{CBRG} \\  \hline \hline
 $E_0$& $1.7754\times 10^{-5}$ & $1.5771\times 10^{-2}$  &
$1.8409\times 10^{-5}$  \\ \hline
$E_1$ & $1.5043\times 10^{-4}$ &
 $4.2679\times 10^{-2} $ & $1.6655\times
10^{-4}$  \\  \hline
 $E_2$ & $4.1761 \times 10^{-4}$ & $4.2794\times 10^{-2}$ &
$4.6408\times 10^{-4}$ \\ \hline
 $E_3$& $8.1831\times 10^{-4}$ & $4.3053\times 10^{-2}$  &
$9.1450\times 10^{-4}$  \\ \hline
 $E_4$& $1.3520\times 10^{-3}$ & $4.3173\times 10^{-2}$  &
$1.5179\times 10^{-3}$  \\ \hline
 $E_5$& $2.0196\times 10^{-3}$ & $4.4288\times 10^{-2}$  &
$2.2852\times 10^{-3}$   \\ \hline \hline
\end{tabular}
\caption{Exact, Standard RG and CBRG Values of  Low Lying
States for the 1D Tight-Binding  Model for a chain of
$N=12 \times 2^5=384$ sites with Free-Fixed BC's.}  \end{table}

\newpage



\begin{table}
\centering
\begin{tabular}{|c|c|c|c|}
\hline \hline
  \mbox{Energies} & \mbox{ Exact}    & \mbox{ Standard BRG}
&  \mbox{CBRG} \\  \hline \hline
 $E_0$& $6.6585\times 10^{-5}$ & $5.8116\times 10^{-2}$  &
$7.0843\times 10^{-5}$  \\ \hline
$E_1$ & $2.6633\times 10^{-4}$ &
 $5.8155\times 10^{-2} $ & $2.9403\times
10^{-4}$  \\  \hline
 $E_2$ & $5.9924\times 10^{-4}$ & $5.8268\times 10^{-2}$ &
$6.3690\times 10^{-4}$ \\ \hline
 $E_3$& $1.0653\times 10^{-3}$ & $5.8470\times 10^{-2}$  &
$1.2289\times 10^{-3}$  \\ \hline
 $E_4$& $1.6644\times 10^{-3}$ & $5.8717\times 10^{-2}$  &
$1.7707\times 10^{-3}$  \\ \hline
 $E_5$& $2.3966\times 10^{-3}$ & $5.9106\times 10^{-2}$  &
$2.7311\times 10^{-3}$   \\ \hline \hline
\end{tabular}
\caption{Exact, Standard RG and CBRG Values of  Low Lying
States for the 1D Tight-Binding  Model for a chain of
$N=12 \times 2^5=384$ sites with Fixed-Fixed BC's.}  \end{table}



\begin{table}
\centering
\begin{tabular}{|c|c|c|}
\hline \hline
  \mbox{Energies} & \mbox{ Exact}
&  \mbox{CBRG} \\  \hline \hline
 $E_0$& $0$   &
$9.6114\times 10^{-35}$  \\ \hline
$E_1$ & $1.5056\times 10^{-4}$  & $1.9390\times
10^{-4}$  \\  \hline
 $E_2$ & $1.5056\times 10^{-4}$  &
$1.9390\times 10^{-4}$ \\  \hline
 $E_3$& $3.0012\times 10^{-4}$  &
$3.8781\times 10^{-4}$  \\ \hline \hline
\end{tabular}
\caption{Exact and CBRG Values of  Low Lying
States for the 2D Tight-Binding  Model for a lattice of
$N=4\times 4 \times 4^6=65536$ sites with Free BC's.}
  \end{table}


\vspace{2cm}









\newpage \begin{thebibliography}{99}


\bibitem{wilson} K.R. Wilson,  {\em  Rev. Mod. Phys.}  {\bf 47}, 773
(1975).


\bibitem{drell} S.D. Drell, M. Weinstein, S. Yankielowicz,
 {\em Phys. Rev.  D} {\bf 16}, 1769 (1977).

\bibitem{jullien} R. Jullien, P. Pfeuty, J.N. Fields, S. Doniach,
{\em  Phys. Rev.
 B} {\bf 18}, 3568 (1978).


\bibitem{white}  S.R. White, {\em Phys. Rev.  Lett.} {\bf 69}, 2863
(1992); {\em Phys. Rev.  B} {\bf 48}, 10345 (1993).

\bibitem{white-noack} S.R. White, R.M. Noack, {\em Phys. Rev.  Lett.
}{\bf 68}, 3487 (1992).

\bibitem{wilsontalk} K.R. Wilson in unpublished talk.


\bibitem{white-huse}  S.R. White and D.A. Huse
 {\em Phys. Rev.  B} {\bf 48}, 3844 (1993).

\bibitem{white-scalapino} R.M.Noack,  S.R. White and D.J. Scalapino
 {\em Phys. Rev.  Lett.} {\bf 73}, 882 (1994).

\bibitem{nishino} T. Nishino and K. Okunishi,
 ``Corner Transfer Matrix Renormalization Group Method"
Preprint 1995 Tohoku University. Cond-mat/9507087.


\bibitem{ostlund-rommer} S. Ostlund  and S. Rommer,
 ``Thermodynamic limit of the density matrix renormalization for
the spin-1 Hesisenberg chain". Preprint March 1995, Goteborg ITP 95-6.

\bibitem{nishino-okunishi} T. Nishino  and K. Okunishi,
 ``Product Wave Function Renormalization Group".
Preprint 1995 Tohoku University. Cond-mat/9510004.

\bibitem {q-germanyo} M.A. Mart\'{\i}n-Delgado and G. Sierra, {\em
1995,  UCM-CSIC preprint},
``Real Space Renormalization Group Methods and
Quantum Groups" to appear in Phys. Rev. Lett.

\bibitem {qbis-germanyo} M.A. Mart\'{\i}n-Delgado and G. Sierra, {\em
1995,  UCM-CSIC preprint},
to appear in ``From Field Theory to Quantum Groups". World Scientific
Publishers.


\bibitem {dm-germanyo} M.A. Mart\'{\i}n-Delgado and G. Sierra, {\em
1995,  UCM-CSIC preprint},
``Analytic Formulations of the Density Matrix Renormalization Group",
to be appear in Int. J. Mod. Phys. {\bf A}.

\bibitem {bc-germanyo} M.A. Mart\'{\i}n-Delgado and G. Sierra, {\em
1995,  UCM-CSIC preprint},
``The Role of Boundary Conditions in the
Real-Space Renormalization Group".
 Phys. Lett. {\bf B364} 41, (1995).


\bibitem {fpacheco}  A. Fern\'andez-Pacheco, {\em Phys. Rev.  D}
{\bf  19}, 3173 (1979).

\bibitem {rabin}J.M. Rabin, {\em Phys. Rev.  B} {\bf 21}, 2027
(1980).

\bibitem {jullienlibro}  Pfeuty, P.Jullien, R. and Penson, K.A., in
``Real-Space Renormalization", editors Burkhardt, T.W. and van
Leeuwen, J.M.J., series topics in Current Physics {\bf 30},
Springer-Verlag 1982.

\bibitem {jaitisi} J. Gonz\'alez, M.A. Mart\'{\i}n-Delgado,
 G. Sierra, A.H. Vozmediano, \\
{\em Quantum Electron Liquids and
High-$T_c$ Superconductivity}, \\
Lecture Notes in Physics, Monographs
vol. {\bf 38}, Springer-Verlag 1995.



\end{thebibliography}
\end{document}